\documentclass[prd,aps,twocolumn,preprintnumbers, showpacs, nofootinbib,superscriptaddress,notitlepage]{revtex4-1}
\usepackage{stmaryrd}
\usepackage{epsfig}
\usepackage{amsfonts}
\usepackage{graphicx}
\usepackage{color}
\usepackage{amssymb,amsmath,psfrag,slashed,graphicx}
\usepackage[normalem]{ulem}
\usepackage[utf8]{inputenc}

\allowdisplaybreaks

\def\nn{\nonumber}

\begin{document}
\title{Weak Decays of Doubly Heavy Baryons: W-Exchange}
\author{Qi-An Zhang~\footnote{Email:zhangqa@ihep.ac.cn}}
\affiliation{Institute of High Energy Physics, Chinese Academy of Science, Beijing 100049, China}
\affiliation{School of Physics, University of Chinese Academy of Sciences, Beijing 100049, China}

\begin{abstract}
Since the LHCb collaboration announced the observation of the doubly charmed baryon $\Xi_{cc}^{++}$, a series of studies of doubly heavy baryons have been presented. In this work, I analyse the non-leptonic weak decays of doubly heavy baryons $\Xi_{bc}$ and $\Omega_{bc}$ under the flavor $SU(3)$ symmetry. I mainly focus on the $W$-exchange diagrams, which will contribute to the decay channels with final states are light meson and light baryon. These channels would be helpful for searching for $\Xi_{bc}$ and $\Omega_{bc}$ at LHC. And these channels and relations of corresponding decay widths could be examined by the future experimental facilities such as LHC, Belle II and CEPC.
\end{abstract}

\maketitle	
	
\section{Introduction}
Quark model \cite{GellMann:1962xb,GellMann:1964nj,Patrignani:2016xqp,Tanabashi:2018oca} has predicted the existence of doubly heavy baryons, while experimentally search for these states has been for a long time \cite{Mattson:2002vu,Ocherashvili:2004hi,Kato:2013ynr,Aaij:2013voa,Aubert:2006qw,Traill:2017zbs,Li:2018epz}. In 2017, the LHCb collaboration has announced the observation of the doubly charmed baryon $\Xi_{cc}^{++}$ with mass $m_{\Xi_{cc}^{++}}=(3621.40\pm0.72\pm0.27\pm0.14)\mathrm{MeV}$ via the weak decay $\Xi_{cc}^{++}\rightarrow\Lambda_c^+K^-\pi^+\pi^+$ \cite{Aaij:2017ueg}. There is no doubt that this landmark discovery will make a substantial impact on both theoretical and experimental sides. Based on this, more experimental observations of the weak decays of doubly heavy baryons are expected, and in the meanwhile, more theoretical and phenomenological efforts are also needed.

The flavor $SU(3)$ symmetry is an effective way to deal with the exclusive decays of hadrons, especially for heavy mesons \cite{Fusheng:2011tw,Li:2013xsa,Zhou:2015jba,Zhou:2016jkv,Wang:2017hxe} or heavy baryons \cite{Li:2017ndo,Lu:2016ogy,Wang:2017azm,Geng:2017mxn,Geng:2018plk,Cheng:2018hwl,Jiang:2018oak,Geng:2018bow,Xing:2018bqt,Zhu:2018epc,Wang:2017mqp,Yu:2017zst,Jiang:2018iqa}, which are difficult to analysed through the factorization theorem (Refs.\cite{Beneke:1999br,Beneke:2003pa,Bauer:2000yr,Keum:2000ph,Keum:2000wi,Lu:2000em,Lu:2000hj,Kurimoto:2001zj} for heavy mesons and Refs.\cite{Wang:2011uv,Mannel:2011xg,Feldmann:2011xf,Ali:2012pn,Detmold:2012vy,Bell:2013tfa,Yu:2017zst} for baryons). This method provides an opportunity to obtain the informations (such as branching fractions, decay widths, etc.) of a series of decay channels once one or a few of them be measured. And if enough experimental data is available, we can extract the $SU(3)$ irreducible amplitudes, and check them with the results calculated in other approaches such like factorization \cite{Wang:2015ndk,Bell:2013tfa,Wang:2009hra,Lu:2009cm,Dhir:2018twm,Lu:2018obb,Zhao:2018zcb,Zhu:2018jet,Sharma:2017txj}. Moreover, we can introduce several nonperturbative parameters to describe the amplitudes and phases in the $SU(3)$ irreducible amplitudes, global fit these parameters by the measured data, and then we can make predictions for decay widths, CP violations of relevant channels. Therefore, we will use the flavor $SU(3)$ symmetry method to investigate the non-leptonic weak decays of doubly heavy baryon $\Xi_{bc}$ and $\Omega_{bc}$.

This work is an extension and supplement of a series of previous work. Ref.\cite{Wang:2017azm} and \cite{Shi:2017dto} have analysed the weak decay of doubly heavy baryons $\Xi_{cc,bc,bb}$ and $\Omega_{cc,bc,bb}$ respectively, and presented the decay amplitudes and relations of decay width for the semi-leptonic and non-leptonic processes. Beyond that, the $W$-exchange diagrams, which are not covered in the previous papers, will be considered in this work. These diagrams are mainly contribute to the channels whose final-state is a light meson and a light baryon. The contributions from the $SU(3)$ singlet $\eta_1$ and $\phi$ are also considered.  In addition, a number of relevant literatures about the weak decays of doubly and triply heavy baryons is listed here: \cite{Wang:2018utj,Chen:2011mb,GomshiNobary:2004mq,Flynn:2011gf,Geng:2017mxn}.

The rest of this paper is organized as follows: A briefly review of the presentations for various hadrons in flavor $SU(3)$ symmetry will be presented in Sec.\ref{sec2}. In Sec.\ref{sec3}, I will analyse the $W$-exchange cases for the $\Xi_{bc}$ and $\Omega_{bc}$ decays, and list the decay amplitudes for each decay modes. In Sec.\ref{sec4}, I will list the golden channels which can be used to discover the corresponding doubly heavy baryons. And a short summary will be presented in the last section.

\section{Particle Multiples}\label{sec2}

In this section, I will begin with a brief review the representations for hadron multiplets under the flavor $SU(3)$ group. Doubly heavy baryons $\Xi_{bc}$ and $\Omega_{bc}$ can form an $SU(3)$ triplet $T_{bc}=\big(\Xi_{bc}^+(bcu),~\Xi_{bc}^0(bcd),~\Omega_{bc}^0(bcs)\big)^T$ with the quantum number of total spin and parity $J^P=1/2^+$.

The light baryons form an $SU(3)$ octet and a decuplet, with the expression of the octet
\begin{align}
T_8= \left(\begin{array}{ccc} \frac{1}{\sqrt{2}}\Sigma^0+\frac{1}{\sqrt{6}}\Lambda^0 & \Sigma^+  &  p  \\ \Sigma^-  &  -\frac{1}{\sqrt{2}}\Sigma^0+\frac{1}{\sqrt{6}}\Lambda^0 & n \\ \Xi^-   & \Xi^0  & -\sqrt{\frac{2}{3}}\Lambda^0
  \end{array} \right),
\end{align}
and the decuplet which is symmetric in $SU(3)$ flavor space:
\begin{align}
&(T_{10})^{111}=\Delta^{++} \nn\\
&(T_{10})^{112}=(T_{10})^{121}=(T_{10})^{211}=\frac{1}{\sqrt3}\Delta^+, \nn\\
&(T_{10})^{222}=\Delta^{-} \nn\\
&(T_{10})^{122}=(T_{10})^{212}=(T_{10})^{221}=\frac{1}{\sqrt3} \Delta^0, \nn\\
&(T_{10})^{113}=(T_{10})^{131}=(T_{10})^{311}=\frac{1}{\sqrt3} \Sigma^{\prime+} \nn\\
&(T_{10})^{223}=(T_{10})^{232}=(T_{10})^{322}= \frac{1}{\sqrt3} \Sigma^{\prime-},\nn\\
&(T_{10})^{123}=(T_{10})^{132}=(T_{10})^{213}=(T_{10})^{231} \nn\\
  &~~~~~~~~~~=(T_{10})^{312}=(T_{10})^{321}= \frac{1}{\sqrt6} \Sigma^{\prime0},\nn\\
&(T_{10})^{133}=(T_{10})^{313}=(T_{10})^{331}= \frac{1}{\sqrt3} \Xi^{\prime0} \nn\\
&(T_{10})^{233}=(T_{10})^{323}=(T_{10})^{332}= \frac{1}{\sqrt3}  \Xi^{\prime-}, \nn\\
&(T_{10})^{333}=\Omega^-.
\end{align}

For the light mesons, pseudoscalar mesons form an $SU(3)$ singlet $M_{1}^{(P)}=\eta_1$ and an octet expressed as
\begin{align}
M_{8}^{(P)}=\begin{pmatrix}
 \frac{1}{\sqrt{2}}\pi^0+\frac{1}{\sqrt{6}}\eta_8   &\pi^+   & K^+\\
 \pi^-  &-\frac{1}{\sqrt{2}}\pi^0+\frac{1}{\sqrt{6}}\eta_8  &{K^0}\\
 K^-  &\bar{K}^0   &-\frac{2}{\sqrt{6}}\eta_8
 \end{pmatrix}.
\end{align}

In particular, it should be noticed that there exists mixing between the flavor eigenstate $(\eta_1,~\eta_8)^T$ and the physical state $(\eta,~\eta^{\prime})^T$ if the flavor $SU(3)$ symmetry is broken, while we do not take this into account in this work.

Similarly, the light vector meson singlet and octet can be written as $M_1^{(V)}=\phi$ and
\begin{align}
M_{8}^{(V)}=\begin{pmatrix}
 \frac{1}{\sqrt{2}}\rho^0+\frac{1}{\sqrt{6}}\omega   &\rho^+   & K^{*+}\\
 \rho^-  &-\frac{1}{\sqrt{2}}\rho^0+\frac{1}{\sqrt{6}}\omega  &{K^{*0}}\\
 K^{*-}  &\bar{K}^{*0}   &-\frac{2}{\sqrt{6}}\omega
 \end{pmatrix}.
\end{align}


\begin{figure*}
\includegraphics[width=1.6\columnwidth]{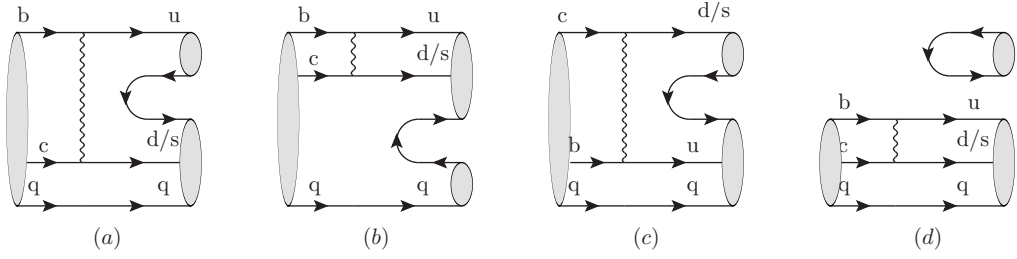}
\caption{Feynman diagrams for $W$-exchange. The first three diagrams exhibit that light meson $SU(3)$ octet in the final state while the last one shows the $SU(3)$ singlet.  If the final state $u$ quark replaced by $c$, these diagrams will contribute to an $SU(3)$ triplet, and have been contained in the $b\rightarrow q_1\bar{q}_2q_3$ cases. So that will not be covered again here.}
\label{fig:W-exchange}
\end{figure*}


\section{Non-leptonic $\Xi_{bc}$ and $\Omega_{bc}$ decays}\label{sec3}

The weak decays of the mixed doubly heavy baryons $\Xi_{bc}$ and $\Omega_{bc}$ can be categorized as three cases: charm and bottom quark decays, and $W$-exchange between the charm and bottom quarks. The quark-level transitions of $b$ and $c$ quark decays are
\begin{align}
&~~~~~c\rightarrow d\bar{d}u,~~~c\rightarrow d\bar{s}u,~~~c\rightarrow s\bar{d}u,~~~c\rightarrow s\bar{s}u, \nn\\
&b\rightarrow c\bar{c}d/s,~~~b\rightarrow c\bar{u}d/s,~~~b\rightarrow u\bar{c}d/s,~~~b\rightarrow q_1\bar{q}_2q_3,
\end{align}
with $q_{1,2,3}$ being light quarks. At hadron level, we can obtain the $\Xi_{bc}$ and $\Omega_{bc}$ decay channels from $\Xi_{cc}$ and $\Omega_{cc}$ decays with the replacement of $T_{cc}\rightarrow T_{bc}$, $T_{c}\rightarrow T_b$ and $D\rightarrow B$, and from $\Xi_{bb}$ and $\Omega_{bb}$ decays with $T_{bb}\rightarrow T_{bc}$, $T_{b}\rightarrow T_c$ and $B\rightarrow D$. Therefore, it is easy to obtain the results for $b$ and $c$ quark decay channels from Ref.\cite{Wang:2017azm} and not necessary to repeat them here.

For the $W$-exchange channels, the effective Hamitonian of the $bc\rightarrow uq$ transition is 
\begin{align}
\mathcal{H}_{eff}~=~&\frac{G_F}{\sqrt{2}}V_{ub}V_{cq}^*[C_1\mathcal{O}_1+C_2\mathcal{O}_2]+h.c.,
\end{align}
where $q=d,s$ and $V_{ub}$, $V_{cq}$ are CKM matrix elements. The transformation of the operators under the flavor $SU(3)$ symmetry as $\bf{3}\otimes\bf{3}=\bf{6}\oplus\bar{\bf{3}}$, so the light quarks form the $\bf{3}$ and $\bf{6}$ representations. The anti-symmetric tensor $H_{\bar{3}}^{\prime\prime}$ and the symmetric tensor $H_6$ have the following nonzero components
\begin{align}
&(H_{\bar{3}}^{\prime\prime})^{12}=-(H_{\bar{3}}^{\prime\prime})^{21}=V_{ub}V_{cd}^*, \\
&(H_6)^{12}=(H_6)^{21}=V_{ub}V_{cd}^*
\end{align}
and 
\begin{align}
&(H_{\bar{3}}^{\prime\prime})^{13}=-(H_{\bar{3}}^{\prime\prime})^{31}=V_{ub}V_{cs}^*, \\
&(H_6)^{13}=(H_6)^{31}=V_{ub}V_{cs}^*
\end{align}
for the transitions $bc\rightarrow ud$ and $bc\rightarrow us$, respectively. Feynman diagrams for these processes are given in Fig.\ref{fig:W-exchange}, one should notice that the final states only contain light quarks. Thus at hadron level, the final state should be a light baryon and a light meson.

\subsection{Decays into a light octet baryon and a light meson}

If the light baryon in the final state is octet, the effective Hamiltonian at hadron level can be constructed as
\begin{align}
\mathcal{H}_{eff}~=~& a_1(T_{bc})^i\epsilon_{ijk}(T_8)_l^k M_n^j(H_{\bar{3}}^{\prime\prime})^{[ln]} \nn\\
	+~& a_2(T_{bc})^i\epsilon_{ijk}(T_8)_l^k M_n^l(H_{\bar{3}}^{\prime\prime})^{[jn]} \nn\\
	+~& a_3(T_{bc})^l\epsilon_{ijk}(T_8)_l^k M_n^i(H_{\bar{3}}^{\prime\prime})^{[jn]} \nn\\
	+~& a_4(T_{bc})^n\epsilon_{ijk}(T_8)_l^k M_n^i(H_{\bar{3}}^{\prime\prime})^{[jl]} \nn\\
	+~& a_5(T_{bc})^n\epsilon_{ijk}(T_8)_l^k M_n^l(H_{\bar{3}}^{\prime\prime})^{[ij]} \nn\\
	+~& a_6(T_{bc})^i\epsilon_{ijk}(T_8)_l^k M_n^j(H_6)^{\{ln\}} \nn\\
	+~& a_7(T_{bc})^i\epsilon_{ijk}(T_8)_l^k M_n^l(H_6)^{\{jn\}} \nn\\
	+~& a_8(T_{bc})^l\epsilon_{ijk}(T_8)_l^k M_n^i(H_6)^{\{jn\}} \nn\\
	+~& a_9(T_{bc})^n\epsilon_{ijk}(T_8)_l^k M_n^i(H_6)^{\{jl\}} \nn\\
	+~&a_{10}(T_{bc})^i\epsilon_{ijk}(T_8)_l^k M_n^n(H_{\bar{3}}^{\prime\prime})^{[jl]}\nn\\
	+~&a_{11}(T_{bc})^i\epsilon_{ijk}(T_8)_l^k M_n^n(H_6)^{\{jl\}}, \label{HamiOfOctet}
\end{align}
in which the coefficients $a_{1\sim11}$ are $SU(3)$ irreducible amplitudes. $a_{1\sim9}$ denote the channels with final state mesons are $SU(3)$ octet, whose Feynman diagrams correspond with the first three diagrams in Fig.\ref{fig:W-exchange}, and $a_{10\sim11}$ denote the $SU(3)$ singlet cases and relevant to the last diagram.

Decay amplitudes for different channels are obtained by expanding the above Hamiltonian and are collected in Tab.\ref{tab:octetBpseuM}. From these amplitudes, we can find the relations for the decay width in the $SU(3)$ limit:
\begin{align}
&\Gamma(\Xi_{bc}^{+}\to\Lambda^0\pi^+)= 2\Gamma(\Xi_{bc}^{0}\to\Lambda^0\pi^0), \nn\\
&\Gamma(\Xi_{bc}^{+}\to\Sigma^0\pi^+ )= \Gamma(\Xi_{bc}^{+}\to\Sigma^+\pi^0 ), \nn\\
&\Gamma(\Omega_{bc}^{0}\to\Sigma^+K^- )= 2\Gamma(\Omega_{bc}^{0}\to\Sigma^0\bar{K}^0 ),\nn\\
&\Gamma(\Omega_{bc}^{0}\to\Xi^-\pi^+ )= 2\Gamma(\Omega_{bc}^{0}\to\Xi^0\pi^0 ), \nn\\
&\Gamma(\Xi_{bc}^{+}\to \Sigma^+\eta_1)=\Gamma(\Omega_{bc}^{0}\to \Xi^0  \eta_1) \nn\\
&\quad\quad\quad\quad\quad\quad~~=-\sqrt{2}\Gamma(\Xi_{bc}^{0}\to \Sigma^0  \eta_1).
\end{align}

It should be noted that above relations for decay widths are only applicable under the flavor $SU(3)$ symmetry, in which the mass differences between final state hadrons have been ignored. And these results will be modified when considering the kinematic corrections of the final phase space.

\small\begin{table*}
\caption{\small{Amplitudes for $\Xi_{bc}$ and $\Omega_{bc}$ decays into a light baryon (octet) and light pseudoscalar meson}}\label{tab:octetBpseuM}
\resizebox{\textwidth}{60mm}{
\scriptsize\begin{tabular}{llll}\hline\hline
channel & amplitude &channel & amplitude \\\hline
$\Xi_{bc}^{+}\to \Lambda^0  \pi^+  $ & $ -\frac{1}{\sqrt{6}}\left(2 a_1-a_2+a_3-a_4+2 a_5+2 a_6+a_7-a_8-3 a_9\right) V_{ub}V_{cs}^*$ &
$\Xi_{bc}^{+}\to \Lambda^0  K^+  $ & $ -\frac{1}{\sqrt{6}}\left(a_1-2 a_2-a_3-2 a_4+4 a_5+a_6+2 a_7+a_8\right) V_{ub}V_{cd}^*$\\
$\Xi_{bc}^{+}\to \Sigma^+  \pi^0  $ & $ \frac{1}{\sqrt{2}}\left(a_2+a_3+a_4-2 a_5-a_7-a_8-a_9\right) V_{ub}V_{cs}^*$ &
$\Xi_{bc}^{+}\to \Sigma^+  K^0  $ & $ \left(a_1+a_3-a_6+a_8\right) V_{ub}V_{cd}^*$\\
$\Xi_{bc}^{+}\to \Sigma^+  \eta_8  $ & $ -\frac{1}{\sqrt{6}}\left(2 a_1-a_2+a_3-a_4+2 a_5-2 a_6+a_7+3 a_8+a_9\right) V_{ub}V_{cs}^*$&
$\Xi_{bc}^{+}\to \Sigma^0  \pi^+  $ & $ -\frac{1}{\sqrt{2}}\left(a_2+a_3+a_4-2 a_5-a_7-a_8-a_9\right) V_{ub}V_{cs}^*$\\
$\Xi_{bc}^{+}\to \Sigma^0  K^+  $ & $ \frac{1}{\sqrt{2}}\left(a_1+a_3+a_6-a_8-2 a_9\right) V_{ub}V_{cd}^*$&
$\Xi_{bc}^{+}\to {p}  \pi^0  $ & $ \frac{1}{\sqrt{2}}\left(a_1-a_2-a_4+2 a_5-a_6+a_7+2 a_8+a_9\right) V_{ub}V_{cd}^*$\\
$\Xi_{bc}^{+}\to {p}  \bar{K}^0  $ & $ -\left(a_1+a_3-a_6+a_8\right) V_{ub}V_{cs}^*$&
$\Xi_{bc}^{+}\to {p}  \eta_8  $ & $ -\frac{1}{\sqrt{6}}\left(a_1+a_2+2 a_3+a_4-2 a_5-a_6-a_7-a_9\right) V_{ub}V_{cd}^*$\\
$\Xi_{bc}^{+}\to {n}  \pi^+  $ & $ \left(a_1-a_2-a_4+2 a_5+a_6+a_7-a_9\right) V_{ub}V_{cd}^*$&
$\Xi_{bc}^{+}\to \Xi^0  K^+  $ & $ -\left(a_1-a_2-a_4+2 a_5+a_6+a_7-a_9\right) V_{ub}V_{cs}^*$\\
$\Xi_{bc}^{0}\to \Lambda^0  \pi^0  $ & $ \frac{1}{2 \sqrt{3}}\left(2 a_1-a_2+a_3-a_4+2 a_5+2 a_6+a_7-a_8-3 a_9\right) V_{ub}V_{cs}^*$&
$\Xi_{bc}^{0}\to \Lambda^0  K^0  $ & $ -\frac{1}{\sqrt{6}}\left(a_1-2 a_2-a_3-2 a_4+4 a_5-a_6-2 a_7-a_8\right) V_{ub}V_{cd}^*$\\
$\Xi_{bc}^{0}\to \Lambda^0  \eta_8  $ & $ \frac{1}{6} \left(4 a_1-5 a_2-a_3+a_4-2 a_5-3 a_7-3 a_8+3 a_9\right) V_{ub}V_{cs}^*$&
$\Xi_{bc}^{0}\to \Sigma^+  \pi^-  $ & $ \left(a_4-2 a_5-a_9\right) V_{ub}V_{cs}^*$\\
$\Xi_{bc}^{0}\to \Sigma^0  \pi^0  $ & $ -\frac{1}{2} \left(a_2+a_3-a_4+2 a_5-a_7-a_8+a_9\right) V_{ub}V_{cs}^*$&
$\Xi_{bc}^{0}\to \Sigma^0  K^0  $ & $ -\frac{1}{\sqrt{2}}\left(a_1+a_3-a_6+a_8+2 a_9\right) V_{ub}V_{cd}^*$\\
$\Xi_{bc}^{0}\to \Sigma^0  \eta_8  $ & $ \frac{1}{2 \sqrt{3}}\left(2 a_1-a_2+a_3-a_4+2 a_5-2 a_6+a_7+3 a_8+a_9\right) V_{ub}V_{cs}^*$&
$\Xi_{bc}^{0}\to \Sigma^-  \pi^+  $ & $ -\left(a_2+a_3-a_7-a_8\right) V_{ub}V_{cs}^*$\\
$\Xi_{bc}^{0}\to \Sigma^-  K^+  $ & $ \left(a_1+a_3+a_6-a_8\right) V_{ub}V_{cd}^*$&
$\Xi_{bc}^{0}\to {p}  \pi^-  $ & $ \left(a_1-a_2-a_4+2 a_5-a_6-a_7+a_9\right) V_{ub}V_{cd}^*$\\
$\Xi_{bc}^{0}\to {p}  K^-  $ & $ \left(a_1-a_2-a_6-a_7\right) V_{ub}V_{cs}^*$&
$\Xi_{bc}^{0}\to {n}  \pi^0  $ & $ -\frac{1}{\sqrt{2}}\left(a_1-a_2-a_4+2 a_5+a_6-a_7-2 a_8-a_9\right) V_{ub}V_{cd}^*$\\
$\Xi_{bc}^{0}\to {n}  \bar{K}^0  $ & $ -\left(a_2+a_3+a_7+a_8\right) V_{ub}V_{cs}^*$&
$\Xi_{bc}^{0}\to {n}  \eta_8  $ & $ -\frac{1}{\sqrt{6}}\left(a_1+a_2+2 a_3+a_4-2 a_5+a_6+a_7+a_9\right) V_{ub}V_{cd}^*$\\	
$\Xi_{bc}^{0}\to \Xi^-  K^+  $ & $ \left(a_1-a_2+a_6+a_7\right) V_{ub}V_{cs}^*$&
$\Xi_{bc}^{0}\to \Xi^0  K^0  $ & $ \left(a_4-2 a_5+a_9\right) V_{ub}V_{cs}^*$\\
$\Omega_{bc}^{0}\to \Lambda^0  \pi^0  $ & $ \frac{1}{\sqrt{3}}\left(a_6-a_7-2 a_8\right) V_{ub}V_{cd}^*$&
$\Omega_{bc}^{0}\to \Lambda^0  \bar{K}^0  $ & $ \frac{1}{\sqrt{6}}\left(a_1+a_2+2 a_3+a_4-2 a_5-a_6+a_7+2 a_8+3 a_9\right) V_{ub}V_{cs}^*$\\
$\Omega_{bc}^{0}\to \Lambda^0  \eta_8  $ & $ \frac{1}{3} \left(a_1+a_2+2 a_3-2 a_4+4 a_5\right) V_{ub}V_{cd}^*$&
$\Omega_{bc}^{0}\to \Sigma^+  \pi^-  $ & $ -\left(a_1-a_2-a_6-a_7\right) V_{ub}V_{cd}^*$\\
$\Omega_{bc}^{0}\to \Sigma^+  K^-  $ & $ -\left(a_1-a_2-a_4+2 a_5-a_6-a_7+a_9\right) V_{ub}V_{cs}^*$&
$\Omega_{bc}^{0}\to \Sigma^0  \pi^0  $ & $ -\left(a_1-a_2\right) V_{ub}V_{cd}^*$\\
$\Omega_{bc}^{0}\to \Sigma^0  \bar{K}^0  $ & $ \frac{1}{\sqrt{2}}\left(a_1-a_2-a_4+2 a_5-a_6-a_7+a_9\right) V_{ub}V_{cs}^*$&
$\Omega_{bc}^{0}\to \Sigma^0  \eta_8  $ & $ -\frac{1}{\sqrt{3}}\left(a_6+a_7-2 a_9\right) V_{ub}V_{cd}^*$\\
$\Omega_{bc}^{0}\to \Sigma^-  \pi^+  $ & $ -\left(a_1-a_2+a_6+a_7\right) V_{ub}V_{cd}^*$&
$\Omega_{bc}^{0}\to {p}  K^-  $ & $ -\left(a_4-2 a_5-a_9\right) V_{ub}V_{cd}^*$\\
$\Omega_{bc}^{0}\to {n}  \bar{K}^0  $ & $ -\left(a_4-2 a_5+a_9\right) V_{ub}V_{cd}^*$&
$\Omega_{bc}^{0}\to \Xi^-  \pi^+  $ & $ -\left(a_1+a_3+a_6-a_8\right) V_{ub}V_{cs}^*$\\
$\Omega_{bc}^{0}\to \Xi^-  K^+  $ & $ \left(a_2+a_3-a_7-a_8\right) V_{ub}V_{cd}^*$&
$\Omega_{bc}^{0}\to \Xi^0  \pi^0  $ & $ \frac{1}{\sqrt{2}}\left(a_1+a_3+a_6-a_8\right) V_{ub}V_{cs}^*$\\
$\Omega_{bc}^{0}\to \Xi^0  K^0  $ & $ \left(a_2+a_3+a_7+a_8\right) V_{ub}V_{cd}^*$&
$\Omega_{bc}^{0}\to \Xi^0  \eta_8  $ & $ \frac{1}{\sqrt{6}}\left(a_1-2 a_2-a_3-2 a_4+4 a_5+a_6-2 a_7-3 a_8-2 a_9\right) V_{ub}V_{cs}^*$\\
$\Xi_{bc}^{+}\to \Sigma^+  \eta_1  $ & $ \left(a_{10}-a_{11}\right) V_{ub}V_{cs}^* $ &
$\Xi_{bc}^{+}\to {p}  \eta_1  $ & $ -\left(a_{10}-a_{11}\right) V_{ub}V_{cd}^*$\\
$\Xi_{bc}^{0}\to \Lambda^0  \eta_1  $ & $ \frac{1}{\sqrt{6}}\left(a_{10}+3 a_{11}\right) V_{ub}V_{cs}^*$ &
$\Xi_{bc}^{0}\to \Sigma^0  \eta_1  $ & $ -\frac{1}{\sqrt{2}}\left(a_{10}-a_{11}\right) V_{ub}V_{cs}^*$\\
$\Xi_{bc}^{0}\to {n}  \eta_1  $ & $ -\left(a_{10}+a_{11}\right) V_{ub}V_{cd}^*$ &
$\Omega_{bc}^{0}\to \Lambda^0  \eta_1  $ & $ \sqrt{\frac{2}{3}} a_{10} V_{ub}V_{cd}^*$\\
$\Omega_{bc}^{0}\to \Sigma^0  \eta_1  $ & $ -\sqrt{2} a_{11} V_{ub}V_{cd}^*$ &
$\Omega_{bc}^{0}\to \Xi^0  \eta_1  $ & $ \left(a_{10}+a_{11}\right) V_{ub}V_{cs}^*$\\
\hline
\end{tabular}}
\end{table*}

Except for the light pseudoscalar meson, we also investigated the light vector meson in the final state. Pseudoscalar meson and vector meson have the same quark constitutions, but with different quantum numbers. They have same effective Hamitonian in Eq.(\ref{HamiOfOctet}), so we can easily get the decay amplitudes for light baryon and vector meson by replacing the final state mesons $\pi\rightarrow\rho$, $K\rightarrow K^*$ and $\eta\rightarrow\omega$. The decay amplitudes for corresponding channels are listed in Tab.(\ref{tab:octetBvectM}), with the following relations for decay width in $SU(3)$ limit:
\begin{align}
&\Gamma(\Xi_{bc}^{+}\to\Lambda^0\rho^+ )= 2\Gamma(\Xi_{bc}^{0}\to\Lambda^0\rho^0 ),\nn\\ 
&\Gamma(\Omega_{bc}^{0}\to\Sigma^+K^{*-} )=2\Gamma(\Omega_{bc}^{0}\to\Sigma^0\bar{K}^{*0} ),\nn\\ 
&\Gamma(\Omega_{bc}^{0}\to\Xi^-\rho^+ )= 2\Gamma(\Omega_{bc}^{0}\to\Xi^0\rho^0 ), \nn\\
&\Gamma(\Xi_{bc}^{+}\to\Sigma^0\rho^+ )= \Gamma(\Xi_{bc}^{+}\to\Sigma^+\rho^0 ),\nn\\
&\Gamma(\Xi_{bc}^{+}\to \Sigma^+\phi)=\Gamma(\Omega_{bc}^{0}\to \Xi^0  \phi) \nn\\
&\quad\quad\quad\quad\quad\quad~~=-\sqrt{2}\Gamma(\Xi_{bc}^{0}\to \Sigma^0  \phi).
\end{align}

\begin{table*}
\caption{Amplitudes for $\Xi_{bc}$ and $\Omega_{bc}$ decays into a light baryon (octet) and light vector meson}\label{tab:octetBvectM}
\resizebox{\textwidth}{60mm}{
\begin{tabular}{llll}\hline\hline
channel & amplitude & channel & amplitude \\\hline
$\Xi_{bc}^{+}\to \Lambda^0  \rho^+  $ & $ -\frac{1}{\sqrt{6}}\left(2 a_1-a_2+a_3-a_4+2 a_5+2 a_6+a_7-a_8-3 a_9\right) V_{ub}V_{cs}^*$&
$\Xi_{bc}^{+}\to \Lambda^0  K^{*+}  $ & $ -\frac{1}{\sqrt{6}}\left(a_1-2 a_2-a_3-2 a_4+4 a_5+a_6+2 a_7+a_8\right) V_{ub}V_{cd}^* $\\
$\Xi_{bc}^{+}\to \Sigma^+  \rho^0  $ & $ \frac{1}{\sqrt{2}}\left(a_2+a_3+a_4-2 a_5-a_7-a_8-a_9\right) V_{ub}V_{cs}^*$&
$\Xi_{bc}^{+}\to \Sigma^+  K^{*0}  $ & $ \left(a_1+a_3-a_6+a_8\right) V_{ub}V_{cd}^* $\\
$\Xi_{bc}^{+}\to \Sigma^+  \omega  $ & $ -\frac{1}{\sqrt{6}}\left(2 a_1-a_2+a_3-a_4+2 a_5-2 a_6+a_7+3 a_8+a_9\right) V_{ub}V_{cs}^*$&
$\Xi_{bc}^{+}\to \Sigma^0  \rho^+  $ & $ -\frac{1}{\sqrt{2}}\left(a_2+a_3+a_4-2 a_5-a_7-a_8-a_9\right) V_{ub}V_{cs}^*$\\
$\Xi_{bc}^{+}\to \Sigma^0  K^{*+}  $ & $ \frac{1}{\sqrt{2}}\left(a_1+a_3+a_6-a_8-2 a_9\right) V_{ub}V_{cd}^* $&
$\Xi_{bc}^{+}\to {p}  \rho^0  $ & $ \frac{1}{\sqrt{2}}\left(a_1-a_2-a_4+2 a_5-a_6+a_7+2 a_8+a_9\right) V_{ub}V_{cd}^* $\\
$\Xi_{bc}^{+}\to {p}  \bar{K}^{*0}  $ & $ -\left(a_1+a_3-a_6+a_8\right) V_{ub}V_{cs}^*$&
$\Xi_{bc}^{+}\to {p}  \omega  $ & $ -\frac{1}{\sqrt{6}}\left(a_1+a_2+2 a_3+a_4-2 a_5-a_6-a_7-a_9\right) V_{ub}V_{cd}^* $\\
$\Xi_{bc}^{+}\to {n}  \rho^+  $ & $ \left(a_1-a_2-a_4+2 a_5+a_6+a_7-a_9\right) V_{ub}V_{cd}^* $&
$\Xi_{bc}^{+}\to \Xi^0  K^{*+}  $ & $ -\left(a_1-a_2-a_4+2 a_5+a_6+a_7-a_9\right) V_{ub}V_{cs}^*$\\
$\Xi_{bc}^{0}\to \Lambda^0  \rho^0  $ & $ \frac{1}{2 \sqrt{3}}\left(2 a_1-a_2+a_3-a_4+2 a_5+2 a_6+a_7-a_8-3 a_9\right) V_{ub}V_{cs}^*$&
$\Xi_{bc}^{0}\to \Lambda^0  K^{*0}  $ & $ -\frac{1}{\sqrt{6}}\left(a_1-2 a_2-a_3-2 a_4+4 a_5-a_6-2 a_7-a_8\right) V_{ub}V_{cd}^* $\\
$\Xi_{bc}^{0}\to \Lambda^0  \omega  $ & $ \frac{1}{6} \left(4 a_1-5 a_2-a_3+a_4-2 a_5-3 a_7-3 a_8+3 a_9\right) V_{ub}V_{cs}^*$&
$\Xi_{bc}^{0}\to \Sigma^+  \rho^-  $ & $ \left(a_4-2 a_5-a_9\right) V_{ub}V_{cs}^*$\\
$\Xi_{bc}^{0}\to \Sigma^0  \rho^0  $ & $ -\frac{1}{2} \left(a_2+a_3-a_4+2 a_5-a_7-a_8+a_9\right) V_{ub}V_{cs}^*$&
$\Xi_{bc}^{0}\to \Sigma^0  K^{*0}  $ & $ -\frac{1}{\sqrt{2}}\left(a_1+a_3-a_6+a_8+2 a_9\right) V_{ub}V_{cd}^* $\\
$\Xi_{bc}^{0}\to \Sigma^0  \omega  $ & $ \frac{1}{2 \sqrt{3}}\left(2 a_1-a_2+a_3-a_4+2 a_5-2 a_6+a_7+3 a_8+a_9\right) V_{ub}V_{cs}^*$&
$\Xi_{bc}^{0}\to \Sigma^-  \rho^+  $ & $ -\left(a_2+a_3-a_7-a_8\right) V_{ub}V_{cs}^*$\\
$\Xi_{bc}^{0}\to \Sigma^-  K^{*+}  $ & $ \left(a_1+a_3+a_6-a_8\right) V_{ub}V_{cd}^* $&
$\Xi_{bc}^{0}\to {p}  \rho^-  $ & $ \left(a_1-a_2-a_4+2 a_5-a_6-a_7+a_9\right) V_{ub}V_{cd}^* $\\
$\Xi_{bc}^{0}\to {p}  K^{*-}  $ & $ \left(a_1-a_2-a_6-a_7\right) V_{ub}V_{cs}^*$&
$\Xi_{bc}^{0}\to {n}  \rho^0  $ & $ -\frac{1}{\sqrt{2}}\left(a_1-a_2-a_4+2 a_5+a_6-a_7-2 a_8-a_9\right) V_{ub}V_{cd}^* $\\
$\Xi_{bc}^{0}\to {n}  \bar{K}^{*0}  $ & $ -\left(a_2+a_3+a_7+a_8\right) V_{ub}V_{cs}^*$&
$\Xi_{bc}^{0}\to {n}  \omega  $ & $ -\frac{1}{\sqrt{6}}\left(a_1+a_2+2 a_3+a_4-2 a_5+a_6+a_7+a_9\right) V_{ub}V_{cd}^* $\\
$\Xi_{bc}^{0}\to \Xi^-  K^{*+}  $ & $ \left(a_1-a_2+a_6+a_7\right) V_{ub}V_{cs}^*$&
$\Xi_{bc}^{0}\to \Xi^0  K^{*0}  $ & $ \left(a_4-2 a_5+a_9\right) V_{ub}V_{cs}^*$\\
$\Omega_{bc}^{0}\to \Lambda^0  \rho^0  $ & $ \frac{1}{\sqrt{3}}\left(a_6-a_7-2 a_8\right) V_{ub}V_{cd}^* $&
$\Omega_{bc}^{0}\to \Lambda^0  \bar{K}^{*0}  $ & $ \frac{1}{\sqrt{6}}\left(a_1+a_2+2 a_3+a_4-2 a_5-a_6+a_7+2 a_8+3 a_9\right) V_{ub}V_{cs}^*$\\
$\Omega_{bc}^{0}\to \Lambda^0  \omega  $ & $ \frac{1}{3} \left(a_1+a_2+2 a_3-2 a_4+4 a_5\right) V_{ub}V_{cd}^* $&
$\Omega_{bc}^{0}\to \Sigma^+  \rho^-  $ & $ -\left(a_1-a_2-a_6-a_7\right) V_{ub}V_{cd}^* $\\
$\Omega_{bc}^{0}\to \Sigma^+  K^{*-}  $ & $ -\left(a_1-a_2-a_4+2 a_5-a_6-a_7+a_9\right) V_{ub}V_{cs}^*$&
$\Omega_{bc}^{0}\to \Sigma^0  \rho^0  $ & $ -\left(a_1-a_2\right) V_{ub}V_{cd}^* $\\
$\Omega_{bc}^{0}\to \Sigma^0  \bar{K}^{*0}  $ & $ \frac{1}{\sqrt{2}}\left(a_1-a_2-a_4+2 a_5-a_6-a_7+a_9\right) V_{ub}V_{cs}^*$&
$\Omega_{bc}^{0}\to \Sigma^0  \omega  $ & $ -\frac{1}{\sqrt{3}}\left(a_6+a_7-2 a_9\right) V_{ub}V_{cd}^* $\\
$\Omega_{bc}^{0}\to \Sigma^-  \rho^+  $ & $ -\left(a_1-a_2+a_6+a_7\right) V_{ub}V_{cd}^* $&
$\Omega_{bc}^{0}\to {p}  K^{*-}  $ & $ -\left(a_4-2 a_5-a_9\right) V_{ub}V_{cd}^* $\\
$\Omega_{bc}^{0}\to {n}  \bar{K}^{*0}  $ & $ -\left(a_4-2 a_5+a_9\right) V_{ub}V_{cd}^* $&
$\Omega_{bc}^{0}\to \Xi^-  \rho^+  $ & $ -\left(a_1+a_3+a_6-a_8\right) V_{ub}V_{cs}^*$\\
$\Omega_{bc}^{0}\to \Xi^-  K^{*+}  $ & $ \left(a_2+a_3-a_7-a_8\right) V_{ub}V_{cd}^* $&
$\Omega_{bc}^{0}\to \Xi^0  \rho^0  $ & $ \frac{1}{\sqrt{2}}\left(a_1+a_3+a_6-a_8\right) V_{ub}V_{cs}^*$\\
$\Omega_{bc}^{0}\to \Xi^0  K^{*0}  $ & $ \left(a_2+a_3+a_7+a_8\right) V_{ub}V_{cd}^* $&
$\Omega_{bc}^{0}\to \Xi^0  \omega  $ & $ \frac{1}{\sqrt{6}}\left(a_1-2 a_2-a_3-2 a_4+4 a_5+a_6-2 a_7-3 a_8-2 a_9\right) V_{ub}V_{cs}^*$\\
$\Xi_{bc}^{+}\to \Sigma^+  \phi  $ & $ \left(a_{10}-a_{11}\right) V_{ub}V_{cs}^* $&
$\Xi_{bc}^{+}\to {p}  \phi  $ & $ -\left(a_{10}-a_{11}\right) V_{ub}V_{cd}^*$\\
$\Xi_{bc}^{0}\to \Lambda^0  \phi  $ & $ \frac{1}{\sqrt{6}}\left(a_{10}+3 a_{11}\right) V_{ub}V_{cs}^*$&
$\Xi_{bc}^{0}\to \Sigma^0  \phi  $ & $ -\frac{1}{\sqrt{2}}\left(a_{10}-a_{11}\right) V_{ub}V_{cs}^*$\\
$\Xi_{bc}^{0}\to {n}  \phi  $ & $ -\left(a_{10}+a_{11}\right) V_{ub}V_{cd}^*$&
$\Omega_{bc}^{0}\to \Lambda^0  \phi  $ & $ \sqrt{\frac{2}{3}} a_{10} V_{ub}V_{cd}^*$\\
$\Omega_{bc}^{0}\to \Sigma^0  \phi  $ & $ -\sqrt{2} a_{11} V_{ub}V_{cd}^*$&
$\Omega_{bc}^{0}\to \Xi^0  \phi  $ & $ \left(a_{10}+a_{11}\right) V_{ub}V_{cs}^*$\\
\hline
\end{tabular}}
\end{table*}

\subsection{Decays into a light decuplet baryon and a light meson}

The effective Hamitonian for the decays of $\Xi_{cc}$ and $\Omega_{cc}$ into a light decuplet baryon and a light meson can be written as
\begin{align}
\mathcal{H}_{eff}~=~& b_1(T_{bc})^i(T_{10})_{ijk}M_l^k(H_{\bar{3}})^{kl} \nn\\
		+~& b_2(T_{bc})^i(T_{10})_{ijk}M_l^k(H_6)^{kl} \nn\\
		+~& b_3(T_{bc})^l(T_{10})_{ijk}M_i^l(H_6)^{jk} \nn\\
		+~& b_4(T_{bc})^i(T_{10})_{ijk}M_l^l(H_6)^{jk} ,
\end{align}
in which $b_{1\sim3}$ denote the cases final state mesons are $SU(3)$ octet while $b_4$ denotes the singlet cases. The corresponding Feynman diagrams have been exhibited in Fig.\ref{fig:W-exchange}.

\begin{table*}
\caption{Amplitudes for $\Xi_{bc}$ and $\Omega_{bc}$ decays into a light baryon (decuplet) and light pseudoscalar meson}\label{tab:decupBpseuM}
\resizebox{\textwidth}{40mm}{
\begin{tabular}{llllll}\hline\hline
channel & amplitude & channel & amplitude & channel & amplitude \\\hline
$\Xi_{bc}^{+}\to \Delta^{++}  \pi^-  $ & $ \left(b_1+b_2\right) V_{ub}V_{cd}^*$ &
$\Xi_{bc}^{+}\to \Delta^{++}  K^-  $ & $ \left(b_1+b_2\right) V_{ub}V_{cs}^*$ &
$\Xi_{bc}^{+}\to \Delta^{+}  \pi^0  $ & $ -\sqrt{\frac{2}{3}} \left(b_1-b_3\right) V_{ub}V_{cd}^*$ \\
$\Xi_{bc}^{+}\to \Delta^{+}  \bar{K}^0  $ & $ \frac{1}{\sqrt{3}}\left(b_1+b_2\right) V_{ub}V_{cs}^*$ &
$\Xi_{bc}^{+}\to \Delta^{+}  \eta_8  $ & $ \frac{1}{3} \sqrt{2} \left(b_2+b_3\right) V_{ub}V_{cd}^*$ &
$\Xi_{bc}^{+}\to \Sigma^{\prime0}  \pi^+  $ & $ -\frac{1}{\sqrt{6}}\left(b_1-b_2-2 b_3\right) V_{ub}V_{cs}^*$\\
$\Xi_{bc}^{+}\to \Delta^{0}  \pi^+  $ & $ -\frac{1}{\sqrt{3}}\left(b_1-b_2-2 b_3\right) V_{ub}V_{cd}^*$ &
$\Xi_{bc}^{+}\to \Sigma^{\prime+}  \pi^0  $ & $ -\frac{1}{\sqrt{6}}\left(b_1-b_2-2 b_3\right) V_{ub}V_{cs}^*$ &
$\Xi_{bc}^{+}\to \Sigma^{\prime+}  K^0  $ & $ \frac{1}{\sqrt{3}}\left(b_1+b_2\right) V_{ub}V_{cd}^*$ \\
$\Xi_{bc}^{+}\to \Sigma^{\prime+}  \eta_8  $ & $ -\frac{1}{3 \sqrt{2}}\left(3 b_1+b_2-2 b_3\right) V_{ub}V_{cs}^*$ &
$\Xi_{bc}^{+}\to \Sigma^{\prime0}  K^+  $ & $ -\frac{1}{\sqrt{6}}\left(b_1-b_2-2 b_3\right) V_{ub}V_{cd}^*$ &
$\Xi_{bc}^{+}\to \Xi^{\prime0}  K^+  $ & $ -\frac{1}{\sqrt{3}}\left(b_1-b_2-2 b_3\right) V_{ub}V_{cs}^*$\\
$\Xi_{bc}^{0}\to \Delta^{+}  \pi^-  $ & $ \frac{1}{\sqrt{3}}\left(b_1+b_2+2 b_3\right) V_{ub}V_{cd}^*$ &
$\Xi_{bc}^{0}\to \Delta^{+}  K^-  $ & $ \frac{1}{\sqrt{3}}\left(b_1+b_2\right) V_{ub}V_{cs}^*$&
$\Xi_{bc}^{0}\to \Delta^{0}  \pi^0  $ & $ -\sqrt{\frac{2}{3}} \left(b_1+b_3\right) V_{ub}V_{cd}^*$ \\
$\Xi_{bc}^{0}\to \Delta^{0}  \bar{K}^0  $ & $ \frac{1}{\sqrt{3}}\left(b_1+b_2\right) V_{ub}V_{cs}^*$&
$\Xi_{bc}^{0}\to \Delta^{0}  \eta_8  $ & $ \frac{1}{3} \sqrt{2} \left(b_2+b_3\right) V_{ub}V_{cd}^*$ &
$\Xi_{bc}^{0}\to \Sigma^{\prime+}  \pi^-  $ & $ \frac{2}{\sqrt{3}} b_3 V_{ub}V_{cs}^*$\\
$\Xi_{bc}^{0}\to \Delta^{-}  \pi^+  $ & $ -\left(b_1-b_2\right) V_{ub}V_{cd}^*$ &
$\Xi_{bc}^{0}\to \Sigma^{\prime0}  \pi^0  $ & $ -\frac{1}{2 \sqrt{3}}\left(b_1-b_2+2 b_3\right) V_{ub}V_{cs}^*$ &
$\Xi_{bc}^{0}\to \Sigma^{\prime0}  K^0  $ & $ \frac{1}{\sqrt{6}}\left(b_1+b_2+2 b_3\right) V_{ub}V_{cd}^*$ \\
$\Xi_{bc}^{0}\to \Sigma^{\prime0}  \eta_8  $ & $ -\frac{1}{6} \left(3 b_1+b_2-2 b_3\right) V_{ub}V_{cs}^*$&
$\Xi_{bc}^{0}\to \Sigma^{\prime-}  K^+  $ & $ -\frac{1}{\sqrt{3}}\left(b_1-b_2\right) V_{ub}V_{cd}^*$ &
$\Xi_{bc}^{0}\to \Sigma^{\prime-}  \pi^+  $ & $ -\frac{1}{\sqrt{3}}\left(b_1-b_2\right) V_{ub}V_{cs}^*$\\
$\Omega_{bc}^{0}\to \Delta^{+}  K^-  $ & $ \frac{2}{\sqrt{3}}b_3 V_{ub}V_{cd}^*$ &
$\Xi_{bc}^{0}\to \Xi^{\prime0}  K^0  $ & $ \frac{2}{\sqrt{3}} b_3 V_{ub}V_{cs}^*$&
$\Omega_{bc}^{0}\to \Delta^{0}  \bar{K}^0  $ & $ \frac{2}{\sqrt{3}} b_3 V_{ub}V_{cd}^*$ \\
$\Xi_{bc}^{0}\to \Xi^{\prime-}  K^+  $ & $ -\frac{1}{\sqrt{3}}\left(b_1-b_2\right) V_{ub}V_{cs}^*$&
$\Omega_{bc}^{0}\to \Sigma^{\prime+}  \pi^-  $ & $ \frac{1}{\sqrt{3}}\left(b_1+b_2\right) V_{ub}V_{cd}^*$ &
$\Omega_{bc}^{0}\to \Sigma^{\prime+}  K^-  $ & $ \frac{1}{\sqrt{3}}\left(b_1+b_2+2 b_3\right) V_{ub}V_{cs}^*$\\
$\Omega_{bc}^{0}\to \Sigma^{\prime0}  \pi^0  $ & $ -\frac{1}{\sqrt{3}}b_1 V_{ub}V_{cd}^*$ &
$\Omega_{bc}^{0}\to \Sigma^{\prime0}  \bar{K}^0  $ & $ \frac{1}{\sqrt{6}}\left(b_1+b_2+2 b_3\right) V_{ub}V_{cs}^*$&
$\Omega_{bc}^{0}\to \Sigma^{\prime0}  \eta_8  $ & $ \frac{1}{3} \left(b_2-2 b_3\right) V_{ub}V_{cd}^*$ \\
$\Omega_{bc}^{0}\to \Xi^{\prime0}  \pi^0  $ & $ -\frac{1}{\sqrt{6}}\left(b_1-b_2\right) V_{ub}V_{cs}^*$&
$\Omega_{bc}^{0}\to \Sigma^{\prime-}  \pi^+  $ & $ -\frac{1}{\sqrt{3}}\left(b_1-b_2\right) V_{ub}V_{cd}^*$ &
$\Omega_{bc}^{0}\to \Xi^{\prime0}  \eta_8  $ & $ -\frac{1}{3 \sqrt{2}}\left(3 b_1+b_2+4 b_3\right) V_{ub}V_{cs}^*$\\
$\Omega_{bc}^{0}\to \Xi^{\prime0}  K^0  $ & $ \frac{1}{\sqrt{3}}\left(b_1+b_2\right) V_{ub}V_{cd}^*$ &
$\Omega_{bc}^{0}\to \Xi^{\prime-}  \pi^+  $ & $ -\frac{1}{\sqrt{3}}\left(b_1-b_2\right) V_{ub}V_{cs}^*$&
$\Omega_{bc}^{0}\to \Xi^{\prime-}  K^+  $ & $ -\frac{1}{\sqrt{3}}\left(b_1-b_2\right) V_{ub}V_{cd}^*$ \\
$\Omega_{bc}^{0}\to \Omega^-  K^+  $ & $ -\left(b_1-b_2\right) V_{ub}V_{cs}^*$ &
$\Xi_{bc}^{+}\to \Delta^{+} \eta_1  $ & $ \frac{2}{\sqrt{3}}b_4 V_{ub}V_{cd}^*$ &
$\Xi_{bc}^{+}\to \Sigma^{\prime+} \eta_1  $ & $ \frac{2}{\sqrt{3}}b_4 V_{ub}V_{cs}^*$\\
$\Xi_{bc}^{0}\to \Delta^{0} \eta_1 $ & $ \frac{2}{\sqrt{3}}b_4 V_{ub}V_{cd}^*$ &
$\Xi_{bc}^{0}\to \Sigma^{\prime0} \eta_1 $ & $ \sqrt{\frac{2}{3}} b_4  V_{ub}V_{cs}^*$ &
$\Omega_{bc}^{0}\to \Sigma^{\prime0} \eta_1 $ & $ \sqrt{\frac{2}{3}} b_4  V_{ub}V_{cd}^*$\\
$\Omega_{bc}^{0}\to \Xi^{\prime0} \eta_1 $ & $ \frac{2}{\sqrt{3}} b_4 V_{ub}V_{cs}^*$\\
\hline
\end{tabular}}
\end{table*}

Expanding the above Hamitonian, we will obtain the decay amplitudes in Tab.\ref{tab:decupBpseuM}, which leads to the following relations of the decay width:
\begin{align}
\Gamma(\Xi_{bc}^{0}\to\Sigma^{\prime-}\pi^+)&=\Gamma(\Xi_{bc}^{0}\to\Xi^{\prime-}K^+ ) 
	 =\Gamma(\Omega_{bc}^{0}\to\Xi^{\prime-}\pi^+ ) \nn\\
	&= 2\Gamma(\Omega_{bc}^{0}\to\Xi^{\prime0}\pi^0 ) 
	 =\frac{1}{3}\Gamma(\Omega_{bc}^{0}\to\Omega^-K^+ ),\nn\\
\Gamma(\Xi_{bc}^{0}\to\Sigma^{\prime-}K^+)&=\Gamma(\Omega_{bc}^{0}\to\Xi^{\prime-}K^+ )
	 =\Gamma(\Omega_{bc}^{0}\to\Sigma^{\prime-}\pi^+ ) \nn\\
	&=\frac{1}{3}\Gamma(\Xi_{bc}^{0}\to\Delta^{-}\pi^+ ),\nn\\
\Gamma(\Xi_{bc}^{+}\to\Sigma^{\prime+}K^0)&=\Gamma(\Omega_{bc}^{0}\to\Xi^{\prime0}K^0 )
	 =\Gamma(\Omega_{bc}^{0}\to\Sigma^{\prime+}\pi^- )\nn\\
	&=\frac{1}{3}\Gamma(\Xi_{bc}^{+}\to\Delta^{++}\pi^- ), \nn\\
\Gamma(\Xi_{bc}^{+}\to\Delta^{+}\bar{K}^0 )&= \Gamma(\Xi_{bc}^{0}\to\Delta^{+}K^- )
	 =\Gamma(\Xi_{bc}^{0}\to\Delta^{0}\bar{K}^0 )\nn\\
	&=\frac{1}{3}\Gamma(\Xi_{bc}^{+}\to\Delta^{++}K^- ),\nn\\
\Gamma(\Xi_{bc}^{+}\to\Sigma^{\prime+}\pi^0 )&=\Gamma(\Xi_{bc}^{+}\to\Sigma^{\prime0}\pi^+ )= \frac{1}{2}\Gamma(\Xi_{bc}^{+}\to\Xi^{\prime0}K^+ ),\nn\\
\Gamma(\Xi_{bc}^{+}\to\Sigma^{\prime0}K^+ )&=\frac{1}{2}\Gamma(\Xi_{bc}^{+}\to\Delta^{0}\pi^+ ),\nn\\
\Gamma(\Xi_{bc}^{0}\to\Sigma^{\prime0}K^0 )&=\frac{1}{2}\Gamma(\Xi_{bc}^{0}\to\Delta^{+}\pi^- ),\nn\\ 
\Gamma(\Omega_{bc}^{0}\to\Sigma^{\prime0}\bar{K}^0 )&= \frac{1}{2}\Gamma(\Omega_{bc}^{0}\to\Sigma^{\prime+}K^- ),\nn\\
\Gamma(\Xi_{bc}^{0}\to\Sigma^{\prime+}\pi^-)&=\Gamma(\Xi_{bc}^{0}\to\Xi^{\prime0}K^0),\nn\\
\Gamma(\Omega_{bc}^{0}\to\Delta^{0}\bar{K}^0)&=\Gamma(\Omega_{bc}^{0}\to\Delta^{+}K^-),\nn\\
\Gamma(\Omega_{bc}^{0}\to\Sigma^{\prime+}\pi^- )&=\Gamma(\Omega_{bc}^{0}\to\Xi^{\prime0}K^0 ), \nn\\
\Gamma(\Xi_{bc}^{+}\to \Delta^{+} \eta_1)&=\Gamma(\Xi_{bc}^{0}\to \Delta^{0} \eta_1)=\sqrt{2}\Gamma(\Omega_{bc}^{0}\to \Sigma^{\prime0} \eta_1),\nn\\
\Gamma(\Xi_{bc}^{+}\to \Sigma^{\prime+} \eta_1)&=\Gamma(\Omega_{bc}^{0}\to \Xi^{\prime0} \eta_1)=\sqrt{2}\Gamma(\Xi_{bc}^{0}\to \Sigma^{\prime0} \eta_1).
\end{align}

And by replacing the pseudoscalar mesons into corresponding vector counterparts, we can list the decay amplitudes for these channels in Tab.(\ref{tab:decupBvecM}) with the following relations for some of the decay widths:
\begin{align}
\Gamma(\Xi_{bc}^{0}\to\Sigma^{\prime-}\rho^+)&=\Gamma(\Xi_{bc}^{0}\to\Xi^{\prime-}K^{*+})
	 =\Gamma(\Omega_{bc}^{0}\to\Xi^{\prime-}\rho^+ ) \nn\\
	&= 2\Gamma(\Omega_{bc}^{0}\to\Xi^{\prime0}\rho^0 ) 
	 =\frac{1}{3}\Gamma(\Omega_{bc}^{0}\to\Omega^-K^{*+} ),\nn\\
\Gamma(\Xi_{bc}^{0}\to\Sigma^{\prime-}K^{*+} )&=\Gamma(\Omega_{bc}^{0}\to\Xi^{\prime-}K^{*+} ) 
	 =\Gamma(\Omega_{bc}^{0}\to\Sigma^{\prime-}\rho^+ )\nn\\
	&=\frac{1}{3}\Gamma(\Xi_{bc}^{0}\to\Delta^{-}\rho^+ ),\nn\\
\Gamma(\Xi_{bc}^{+}\to\Sigma^{\prime+}K^{*0} )&=\Gamma(\Omega_{bc}^{0}\to\Xi^{\prime0}K^{*0} )=\Gamma(\Omega_{bc}^{0}\to\Sigma^{\prime+}\rho^- ) \nn\\
	&=\frac{1}{3}\Gamma(\Xi_{bc}^{+}\to\Delta^{++}\rho^- ), \nn\\
\Gamma(\Xi_{bc}^{+}\to\Delta^{+}\bar{K}^{*0} )&= \Gamma(\Xi_{bc}^{0}\to\Delta^{+}K^{*-} )=\Gamma(\Xi_{bc}^{0}\to\Delta^{0}\bar{K}^{*0} ) \nn\\
	&=\frac{1}{3}\Gamma(\Xi_{bc}^{+}\to\Delta^{++}K^{*-} ),\nn\\
\Gamma(\Xi_{bc}^{+}\to\Sigma^{\prime+}\rho^0 )&=\Gamma(\Xi_{bc}^{+}\to\Sigma^{\prime0}\rho^+ )= \frac{1}{2}\Gamma(\Xi_{bc}^{+}\to\Xi^{\prime0}K^{*+} ),\nn\\
\Gamma(\Xi_{bc}^{+}\to\Sigma^{\prime0}K^{*+} )&=\frac{1}{2}\Gamma(\Xi_{bc}^{+}\to\Delta^{0}\rho^+ ),\nn\\
\Gamma(\Xi_{bc}^{0}\to\Sigma^{\prime0}K^{*0} )&=\frac{1}{2}\Gamma(\Xi_{bc}^{0}\to\Delta^{+}\rho^- ),\nn\\ 
\Gamma(\Omega_{bc}^{0}\to\Sigma^{\prime0}\bar{K}^{*0} )&= \frac{1}{2}\Gamma(\Omega_{bc}^{0}\to\Sigma^{\prime+}K^{*-} ),\nn\\
\Gamma(\Xi_{bc}^{0}\to\Sigma^{\prime+}\rho^- )&=\Gamma(\Xi_{bc}^{0}\to\Xi^{\prime0}K^{*0} ),\nn\\
\Gamma(\Omega_{bc}^{0}\to\Delta^{0}\bar{K}^{*0} )&=\Gamma(\Omega_{bc}^{0}\to\Delta^{+}K^{*-} ),\nn\\
\Gamma(\Omega_{bc}^{0}\to\Sigma^{\prime+}\rho^- )&=\Gamma(\Omega_{bc}^{0}\to\Xi^{\prime0}K^{*0} ),\nn\\
\Gamma(\Xi_{bc}^{+}\to \Delta^{+} \phi)&=\Gamma(\Xi_{bc}^{0}\to \Delta^{0} \phi)=\sqrt{2}\Gamma(\Omega_{bc}^{0}\to \Sigma^{\prime0} \phi),\nn\\
\Gamma(\Xi_{bc}^{+}\to \Sigma^{\prime+} \phi)&=\Gamma(\Omega_{bc}^{0}\to \Xi^{\prime0} \phi)=\sqrt{2}\Gamma(\Xi_{bc}^{0}\to \Sigma^{\prime0} \phi).
\end{align}

\begin{table*}
\caption{Amplitudes for $\Xi_{bc}$ and $\Omega_{bc}$ decays into a light baryon (decuplet) and light vector meson}\label{tab:decupBvecM}
\resizebox{\textwidth}{40mm}{
\begin{tabular}{llllll}\hline\hline
channel & amplitude & channel & amplitude & channel & amplitude \\\hline
$\Xi_{bc}^{+}\to \Delta^{++}  \rho^-  $ & $ \left(b_1+b_2\right) V_{ub}V_{cd}^*$ &
$\Xi_{bc}^{+}\to \Delta^{++}  K^{*-}  $ & $ \left(b_1+b_2\right) V_{ub}V_{cs}^*$ &
$\Xi_{bc}^{+}\to \Delta^{+}  \rho^0  $ & $ -\sqrt{\frac{2}{3}} \left(b_1-b_3\right) V_{ub}V_{cd}^*$ \\
$\Xi_{bc}^{+}\to \Delta^{+}  \bar{K}^{*0}  $ & $ \frac{1}{\sqrt{3}}\left(b_1+b_2\right) V_{ub}V_{cs}^*$ &
$\Xi_{bc}^{+}\to \Delta^{+}  \omega  $ & $ \frac{1}{3} \sqrt{2} \left(b_2+b_3\right) V_{ub}V_{cd}^*$ &
$\Xi_{bc}^{+}\to \Sigma^{\prime0}  \rho^+  $ & $ -\frac{1}{\sqrt{6}}\left(b_1-b_2-2 b_3\right) V_{ub}V_{cs}^*$ \\
$\Xi_{bc}^{+}\to \Delta^{0}  \rho^+  $ & $ -\frac{1}{\sqrt{3}}\left(b_1-b_2-2 b_3\right) V_{ub}V_{cd}^*$ &
$\Xi_{bc}^{+}\to \Sigma^{\prime+}  \rho^0  $ & $ -\frac{1}{\sqrt{6}}\left(b_1-b_2-2 b_3\right) V_{ub}V_{cs}^*$ &
$\Xi_{bc}^{+}\to \Sigma^{\prime+}  K^{*0}  $ & $ \frac{1}{\sqrt{3}}\left(b_1+b_2\right) V_{ub}V_{cd}^*$ \\
$\Xi_{bc}^{+}\to \Sigma^{\prime+}  \omega  $ & $ -\frac{1}{3 \sqrt{2}}\left(3 b_1+b_2-2 b_3\right) V_{ub}V_{cs}^*$ &
$\Xi_{bc}^{+}\to \Sigma^{\prime0}  K^{*+}  $ & $ -\frac{1}{\sqrt{6}}\left(b_1-b_2-2 b_3\right) V_{ub}V_{cd}^*$ &
$\Xi_{bc}^{+}\to \Xi^{\prime0}  K^{*+}  $ & $ -\frac{1}{\sqrt{3}}\left(b_1-b_2-2 b_3\right) V_{ub}V_{cs}^*$ \\
$\Xi_{bc}^{0}\to \Delta^{+}  \rho^-  $ & $ \frac{1}{\sqrt{3}}\left(b_1+b_2+2 b_3\right) V_{ub}V_{cd}^*$ &
$\Xi_{bc}^{0}\to \Delta^{+}  K^{*-}  $ & $ \frac{1}{\sqrt{3}}\left(b_1+b_2\right) V_{ub}V_{cs}^*$ &
$\Xi_{bc}^{0}\to \Delta^{0}  \rho^0  $ & $ -\sqrt{\frac{2}{3}} \left(b_1+b_3\right) V_{ub}V_{cd}^*$ \\
$\Xi_{bc}^{0}\to \Delta^{0}  \bar{K}^{*0}  $ & $ \frac{1}{\sqrt{3}}\left(b_1+b_2\right) V_{ub}V_{cs}^*$ &
$\Xi_{bc}^{0}\to \Delta^{0}  \omega  $ & $ \frac{1}{3} \sqrt{2} \left(b_2+b_3\right) V_{ub}V_{cd}^*$ &
$\Xi_{bc}^{0}\to \Sigma^{\prime+}  \rho^-  $ & $ \frac{2}{\sqrt{3}} b_3 V_{ub}V_{cs}^*$\\
$\Xi_{bc}^{0}\to \Delta^{-}  \rho^+  $ & $ -\left(b_1-b_2\right) V_{ub}V_{cd}^*$ &
$\Xi_{bc}^{0}\to \Sigma^{\prime0}  \rho^0  $ & $ -\frac{1}{2 \sqrt{3}}\left(b_1-b_2+2 b_3\right) V_{ub}V_{cs}^*$ &
$\Xi_{bc}^{0}\to \Sigma^{\prime0}  K^{*0}  $ & $ \frac{1}{\sqrt{6}}\left(b_1+b_2+2 b_3\right) V_{ub}V_{cd}^*$ \\
$\Xi_{bc}^{0}\to \Sigma^{\prime0}  \omega  $ & $ -\frac{1}{6} \left(3 b_1+b_2-2 b_3\right) V_{ub}V_{cs}^*$ &
$\Xi_{bc}^{0}\to \Sigma^{\prime-}  K^{*+}  $ & $ -\frac{1}{\sqrt{3}}\left(b_1-b_2\right) V_{ub}V_{cd}^*$ &
$\Xi_{bc}^{0}\to \Sigma^{\prime-}  \rho^+  $ & $ -\frac{1}{\sqrt{3}}\left(b_1-b_2\right) V_{ub}V_{cs}^*$ \\
$\Omega_{bc}^{0}\to \Delta^{+}  K^{*-}  $ & $ \frac{2}{\sqrt{3}}b_3 V_{ub}V_{cd}^*$ &
$\Xi_{bc}^{0}\to \Xi^{\prime0}  K^{*0}  $ & $ \frac{2}{\sqrt{3}} b_3 V_{ub}V_{cs}^*$ &
$\Omega_{bc}^{0}\to \Delta^{0}  \bar{K}^{*0}  $ & $ \frac{2}{\sqrt{3}} b_3 V_{ub}V_{cd}^*$\\
$\Xi_{bc}^{0}\to \Xi^{\prime-}  K^{*+}  $ & $ -\frac{1}{\sqrt{3}}\left(b_1-b_2\right) V_{ub}V_{cs}^*$ &
$\Omega_{bc}^{0}\to \Sigma^{\prime+}  \rho^-  $ & $ \frac{1}{\sqrt{3}}\left(b_1+b_2\right) V_{ub}V_{cd}^*$ &
$\Omega_{bc}^{0}\to \Sigma^{\prime+}  K^{*-}  $ & $ \frac{1}{\sqrt{3}}\left(b_1+b_2+2 b_3\right) V_{ub}V_{cs}^*$ \\
$\Omega_{bc}^{0}\to \Sigma^{\prime0}  \rho^0  $ & $ -\frac{1}{\sqrt{3}}b_1 V_{ub}V_{cd}^*$ &
$\Omega_{bc}^{0}\to \Sigma^{\prime0}  \bar{K}^{*0}  $ & $ \frac{1}{\sqrt{6}}\left(b_1+b_2+2 b_3\right) V_{ub}V_{cs}^*$ &
$\Omega_{bc}^{0}\to \Sigma^{\prime0}  \omega  $ & $ \frac{1}{3} \left(b_2-2 b_3\right) V_{ub}V_{cd}^*$ \\
$\Omega_{bc}^{0}\to \Xi^{\prime0}  \rho^0  $ & $ -\frac{1}{\sqrt{6}}\left(b_1-b_2\right) V_{ub}V_{cs}^*$&
$\Omega_{bc}^{0}\to \Sigma^{\prime-}  \rho^+  $ & $ -\frac{1}{\sqrt{3}}\left(b_1-b_2\right) V_{ub}V_{cd}^*$ &
$\Omega_{bc}^{0}\to \Xi^{\prime0}  \omega  $ & $ -\frac{1}{3 \sqrt{2}}\left(3 b_1+b_2+4 b_3\right) V_{ub}V_{cs}^*$ \\
$\Omega_{bc}^{0}\to \Xi^{\prime0}  K^{*0}  $ & $ \frac{1}{\sqrt{3}}\left(b_1+b_2\right) V_{ub}V_{cd}^*$ &
$\Omega_{bc}^{0}\to \Xi^{\prime-}  \rho^+  $ & $ -\frac{1}{\sqrt{3}}\left(b_1-b_2\right) V_{ub}V_{cs}^*$ &
$\Omega_{bc}^{0}\to \Xi^{\prime-}  K^{*+}  $ & $ -\frac{1}{\sqrt{3}}\left(b_1-b_2\right) V_{ub}V_{cd}^*$ \\
$\Omega_{bc}^{0}\to \Omega^-  K^{*+}  $ & $ -\left(b_1-b_2\right) V_{ub}V_{cs}^*$ &
$\Xi_{bc}^{+}\to \Delta^{+} \phi  $ & $ \frac{2}{\sqrt{3}}b_4 V_{ub}V_{cd}^*$ &
$\Xi_{bc}^{+}\to \Sigma^{\prime+} \phi  $ & $ \frac{2}{\sqrt{3}}b_4 V_{ub}V_{cs}^*$\\
$\Xi_{bc}^{0}\to \Delta^{0} \phi $ & $ \frac{2}{\sqrt{3}}b_4 V_{ub}V_{cd}^*$ &
$\Xi_{bc}^{0}\to \Sigma^{\prime0} \phi $ & $ \sqrt{\frac{2}{3}} b_4  V_{ub}V_{cs}^*$ &
$\Omega_{bc}^{0}\to \Sigma^{\prime0} \phi $ & $ \sqrt{\frac{2}{3}} b_4  V_{ub}V_{cd}^*$\\
$\Omega_{bc}^{0}\to \Xi^{\prime0} \phi $ & $ \frac{2}{\sqrt{3}} b_4 V_{ub}V_{cs}^*$\\
\hline
\end{tabular}}
\end{table*}

\section{Golden Decay Channels}\label{sec4}

Based on the previous results of decay amplitudes and relations of decay width, we will list the golden channels in this section, according to the following conditions: 	

\begin{itemize}
\item CKM matrix elements: Note that the decay amplitudes in previous analysis contain a overall CKM factor $V_{ub}V_{cd}^*$ or $V_{ub}V_{cs}^*$, in which the matrix element $V_{cd}\sim0.2$ is much smaller than $V_{cs}\sim1$ \cite{Patrignani:2016xqp,Tanabashi:2018oca}. Only the CKM allowed decay channels for $\Xi_{bc}$ and $\Omega_{bc}$ will be considered.

\item Mesons: In terms of the experimental observations, the final neutral mesons, such like $\pi^0$, $\eta$, $\phi$, $\rho^{\pm}$ (decay into $\pi^{\pm}\pi^0$), $K^{*\pm}$ (decay into $K^{\pm}\pi^0$) and $\omega$ (decays into $\pi^+\pi^-\pi^0$) are difficult to reconstruct at LHC, so the decay modes related with these final state mesons will be ignored. 

\item Baryons: For reconstructing the final state baryons, the primary decay modes of them with corresponding fraction are listed in Tab.\ref{tab:DecayModesOfLightBaryons} \cite{Patrignani:2016xqp,Tanabashi:2018oca}. Similarly to above, by throwing away whose final state contains neutral particles, $\Lambda^0$, $\Sigma^0$, $\Xi^-$, $\Delta^{++}$, $\Delta^0$, $\Sigma^{\prime\pm}$ and $\Xi^{\prime0}$ are all optional candidates.

\item Detection efficiency: By considering the particle detectors in LHC, the detection efficiency of proton is higher than charged pion, and charged pion is higher than photon. And for the cascade decays, with the numbers of final state particles increasing, the detection efficiency of the detectors will necessarily decline. Moreover, the multibody phase space will be suppressed with the increase of final state particles.

\end{itemize}

\begin{table}
\caption{\small{Primary decay modes of light baryons}}\label{tab:DecayModesOfLightBaryons}
\begin{tabular}{llllllll}\hline\hline
Decay modes (fractions) \quad\quad\quad\quad\quad\quad &   \\\hline
$\Lambda^0\rightarrow p\pi^-$ ($\sim63.9\% $)  & $\surd$ \\
$\Sigma^+\rightarrow p\pi^0/n\pi^+$ & $\times$ \\
$\Sigma^+\rightarrow n\pi^- $ & $\times$ \\
$\Sigma^0\rightarrow\Lambda^0\gamma$ ($\sim 100\%$) & suppressed\\
$\Xi^0\rightarrow\Lambda^0\pi^0$ & $\times$ \\
$\Xi^-\rightarrow\Lambda^0\pi^-$ ($\sim99.9\%$) & suppressed\\
$\Delta^{++}\rightarrow p\pi^+$ ($\sim99.4\%$) & $\surd$ \\
$\Delta^+\rightarrow p\pi^0/n\pi^+$ & $\times$ \\
$\Delta^0\rightarrow p\pi^-$ ($\sim99.4\%$) & $\surd$ \\
$\Sigma^{\prime\pm}\rightarrow\Lambda^0\pi^{\pm}$ ($\sim87.0\%$) & suppressed \\
$\Xi^{\prime0}\rightarrow\Xi^0\pi^0(\times)/\Xi^-\pi^+$  & suppressed\\
$\Xi^{\prime-}\rightarrow\Xi^-\pi^0/\Xi^0\pi^-$  &  $\times$ \\
$\Omega^-\rightarrow\Lambda^0K^-$ ($\sim67.8\%$) & suppressed \\
\hline
\end{tabular}
\end{table}

To sum up, by considering the above factors, we list the all allowed decay channels of $\Xi_{bc}$ and $\Omega_{bc}$ (include the suppressed ones) in Tab.\ref{tab:AllowedChannels}. Among these channels, $\Xi_{bc}\rightarrow pK$ or $\Xi_{bc}\rightarrow \Delta K\rightarrow p\pi K$ might be used for searching $\Xi_{bc}$, and $\Omega_{bc}\rightarrow\Lambda^0K$ might be used to search for $\Omega_{bc}$ at LHC.

\begin{table}
\caption{\small{Allowed decay channels of $\Xi_{bc}$ and $\Omega_{bc}$}}\label{tab:AllowedChannels}
\scriptsize\begin{tabular}{llllllll}\hline\hline
channel  \\\hline
$\Xi_{bc}^{+}\to \Lambda^0  \pi^+  $ &
$\Xi_{bc}^{+}\to \Sigma^0  \pi^+  $ &
$\Xi_{bc}^{+}\to {p}  \bar{K}^0  $ &
$\Xi_{bc}^{0}\to {p}  K^-  $ &  \\
$\Xi_{bc}^{0}\to \Xi^-  K^+  $ & 
$\Xi_{bc}^{+}\to {p}  \bar{K}^{*0}  $ & 
$\Xi_{bc}^{0}\to \Lambda^0  \rho^0  $ & 
$\Xi_{bc}^{0}\to \Sigma^0  \rho^0  $ & \\
$\Xi_{bc}^{+}\to \Delta^{++}  K^-  $ & 
$\Xi_{bc}^{+}\to \Xi^{\prime0}  K^+  $ & 
$\Xi_{bc}^{0}\to \Delta^{0}  \bar{K}^0  $ & 
$\Xi_{bc}^{0}\to \Sigma^{\prime+}  \pi^-  $ & \\
$\Xi_{bc}^{0}\to \Sigma^{\prime-}  \pi^+  $ & 
$\Xi_{bc}^{0}\to \Xi^{\prime0}  K^0  $ & 
$\Xi_{bc}^{+}\to \Sigma^{\prime+}  \rho^0  $ & 
$\Xi_{bc}^{0}\to \Delta^{0}  \bar{K}^{*0}  $ & \\
$\Xi_{bc}^{0}\to \Xi^{\prime0}  K^{*0}  $ & \\
\hline
$\Omega_{bc}^{0}\to \Lambda^0  \bar{K}^0  $ & 
$\Omega_{bc}^{0}\to \Sigma^0  \bar{K}^0  $ &
$\Omega_{bc}^{0}\to \Xi^-  \pi^+  $ & 
$\Omega_{bc}^{0}\to \Lambda^0  \bar{K}^{*0}  $ &\\
$\Omega_{bc}^{0}\to \Sigma^0  \bar{K}^{*0}  $ & 
$\Omega_{bc}^{0}\to \Sigma^{\prime+}  K^-  $ & 
$\Omega_{bc}^{0}\to \Omega^-  K^+  $ & 
$\Omega_{bc}^{0}\to \Xi^{\prime0}  \rho^0  $ & \\
\hline
\end{tabular}
\end{table}

\section{Conclusions}

In this work, I analysed the non-leptonic weak decays of doubly heavy baryons $\Xi_{bc}$ and $\Omega_{bc}$ under the flavor $SU(3)$ symmetry. Based on a series of previous achievements, I studied the $W$-exchange diagrams, which mainly contribute to the decay modes of final states are light baryon and light meson. These modes would be helpful to search for $\Xi_{bc}$ and $\Omega_{bc}$ in the experiments. In the flavor $SU(3)$ symmetry approach, decay amplitudes for various decay channels can be parametrized as the sum of a series of $SU(3)$ irreducible amplitudes. And based on this, a number of relations or sum rules between the decay width have been presented, which can be examined by the future experimental measurements in LHC, Belle II or CEPC. Several decay channels have been listed such like $\Xi_{bc}\rightarrow pK$, $\Xi_{bc}\rightarrow \Delta K\rightarrow p\pi K$ and $\Omega_{bc}\rightarrow\Lambda^0K$, which might be helpful to search for $\Xi_{bc}$ and $\Omega_{bc}$ especially at LHC, since their branching fractions are sizeable, and the productions are easily to identify.

\section*{Acknowledgement}

I am grateful to Pro. Cai-Dian Lu, Prof. Wei Wang and Ji Xu for the inspirations and helpful discussions, and thank the hospitality from Prof. Wei Wang at Shanghai Jiao Tong University, where this work was accomplished. This work is supported by National Natural Science Foundation of China under Grant No.11621131001 and 11521505.


\end{document}